\documentstyle[aas2pp4]{article}
\lefthead{Monnier et al.}
\righthead{Non-Uniform Outflow around NML Cyg}

\begin{document}
%_____________________________________TITLE PAGE___________________________
\title{Non-Uniform Dust Outflow Observed\\
around Infrared Object NML Cygni}
\author{J. D. Monnier,  M. Bester, W. C. Danchi, M. A. Johnson\altaffilmark{1},
E. A. Lipman, C. H. Townes, and P.~G.~Tuthill
} 
\affil{Space Sciences Laboratory, University of California, Berkeley,
Berkeley,  CA  94720-7450, USA
}
\author{T. R. Geballe}
\affil{Joint Astronomy Centre,
600 North A'ohoku Place, University Park, Hilo, HI 96720, USA}
\altaffiltext{1}{on leave from Lawrence Livermore National Laboratory, 
Livermore, CA, 94550, USA}
\author{D. Nishimoto}
\affil{Rockwell Power Systems,
535 Lipoa Parkway, Suite 200, 
Kihei, HI 96753, USA}
\author{P. W. Kervin}
\affil{USAF Phillips Laboratory, 
535 Lipoa Parkway, 
Kihei, HI 96753, USA}
%
%
%____________________________________ABSTRACT PAGE_________________________
\begin{abstract}

Measurements by the U.C. Berkeley Infrared Spatial Interferometer 
at 11.15~$\micron$ have yielded strong evidence for multiple dust
shells and/or significant asymmetric dust emission around NML~Cyg.  New
observations reported also include multiple 8-13~$\micron$ spectra taken
from 1994-1995 and N band (10.2~$\micron$) photometry from 1980-1992.
These and past measurements are analyzed and fitted to a model of the
dust distribution around NML~Cyg.  No spherically symmetric 
single dust shell model is found consistent with both
near- and mid-infrared observations.  However, a circularly symmetric
maximum entropy reconstruction of the 11~$\micron$ brightness
distribution suggests a double shell model for the dust distribution.
Such a model, consisting of a geometrically thin shell of intermediate
optical depth ($\tau_{11\mu \rm{m}} \sim 1.9$) plus an outer shell
($\tau_{11\mu \rm{m}} \sim 0.33$), is consistent not only
with the 11~$\micron$ visibility data, but also with near-infrared
speckle measurements, the broadband spectrum, and the 9.7~$\micron$
silicate feature.  The outer shell, or large scale structure, is
revealed only by long-baseline interferometry at 11~$\micron$, being
too cold ($\sim$400~K) to contribute in the near-infrared and having no
unambiguous spectral signature in the mid-infrared. The optical constants
of Ossenkopf, Henning, \& Mathis (1992) proved superior to the
Draine \& Lee (1984) constants in fitting the detailed shape of the 
silicate feature and broadband spectrum for this object.  Recent
observations of H$_2$O maser emission around NML~Cyg by Richards,
Yates, \& Cohen (1996) are consistent with the location of the two dust
shells and provide further evidence for the two-shell model.  

\end{abstract}

%\keywords{asymptotic giant branch stars, dust optical constants, 
%variable stars, dust formation, mass loss, infrared interferometry, 
%stars-individual: NML~Cyg}

%_______________________________________INTRODUCTION_______________________
\section{Introduction}

NML~Cyg is believed to be a supergiant surrounded by an optically
thick, dusty envelope.  An oxygen-rich atmosphere is evidenced in the
mid-infrared spectrum by the presence of a silicate feature in partial
absorption. The high optical depth of the enshrouding material is
clearly shown by the extreme redness of the broadband spectrum.  This
spectrum has been fitted by Rowan-Robinson \& Harris (1983) by modeling
the circumstellar environment as a single shell of dirty silicate dust,
assuming a spherically symmetric, uniform outflow.  A uniform outflow would
produce a $\rho_{\rm{dust}}\propto r^{-2}$ dust density distribution, assuming
the dust is accelerated to its terminal velocity near the star.  
Observations of a dust distribution not consistent with 
$\rho_{\rm{dust}}\propto r^{-2}$ would imply the mass loss is episodic
and may reveal new information about the physical processes involved,
such as the relevant time and length scales.

A target of early workers in speckle interferometry, NML~Cyg has had
its angular size measured at various infrared wavelengths (Sibille,
Chelli, \& Lena 1979; Dyck et al. 1984; Ridgway et al. 1986; Fix \&
Cobb 1988; Dyck \& Benson 1992). Ridgway et al. (1986) demonstrated
that the near-infrared visibility measurements can be explained by a
dust shell geometry similar to that in the uniform outflow model of
Rowan-Robinson \& Harris (1983), however this new model did require the
introduction of excess stellar flux, presumably escaping through less
dense regions in the dust envelope or by excess forward-scattering off
of grains.

Other workers (Rowan-Robinson 1982; Rowan-Robinson \& Harris 1983;
David \& Papoular 1990, 1992), while finding uniform outflow models to
be consistent with giants whose silicate feature is in emission where
the dust is optically thin, have
found indications of other dust shell geometries for optically thick
envelopes.  NML~Cyg belongs to neither group, its silicate feature not
evincing strong emission nor deep absorption.  Dynamical simulations of
the extended envelopes of carbon stars by Winters et al. (1994, 1995)
and Fleischer, Gauger, \& Sedlmayr (1995) show multiple shell
structures caused by an interplay of the dust formation process and the
underlying stellar pulsation.  Such structures are also expected to
arise in the atmospheres of oxygen-rich stars.

This paper introduces new 11.15~$\micron$ observations of the dust
shell around NML~Cyg with the U.C. Berkeley Infrared Spatial
Interferometer (ISI), recent measurements of the mid-infrared spectrum
using the United Kingdom Infrared Telescope, and systematic long-term N
Band (10.2~$\micron$) photometry at the Maui Space Surveillance Site.
Radiative transfer calculations were performed to determine whether
previously published models of the circumstellar dust shell were
consistent with these new data.  We were unable to produce a
uniform outflow model which could fit both the new and previously
published observations; however, a double shell model
satisfactorily does so.

Uniform outflow and spherical symmetry are usually assumed in the 
modeling of evolved stars since the lack of high resolution imaging data
does not require a more detailed model (see Lopez et al. [1996] for recent 
modeling efforts not assuming
spherical symmetry).  These simplifications will come under greater
scrutiny as the dust envelopes are imaged at successively higher
resolution.  Given the known gas velocities ($\sim$5-30~km/s) and
estimated distances of close red giants ($\sim$100-200~pc), dust will
move across one ISI ``resolution element'' of 30~mas in a time scale of 
order the
star's pulsational period ($\sim$400~days).  This suggests that any
pulsation-related inhomogeneities associated with the mass loss
mechanisms can be observed by the ISI, assuming velocity gradients are
not large enough to disperse the dust into an an average $r^{-2}$
density distribution.  With a CO outflow velocity measured to be
$\sim$25~km/s (Knapp et al. 1982; Bowers, Johnston, \& Spencer 1983;
Morris \& Jura 1983), NML~Cyg's greater distance of approximately
1800~pc (Bowers et al. 1983; Morris \& Jura 1983; Richards et al. 1996)
and longer pulsational period of $\sim$940~days as determined from
N~band photometry (Figure~1) imply that the highest resolution spatial
structure observed by the ISI corresponds to a temporal scale of
$\sim$10~yrs ($\sim$3~cycles), making observations of pulsation-related
dust inhomogeneities possible.

%____________________________PRESENT_OBSERVATIONS_______________________
\section{Observations}

\subsection{Mid-Infrared Visibility Measurements} 

NML~Cyg was observed in July and September of both 1993 and 1994 and in 
September and October of 1996 by the
U.C.  Berkeley Infrared Spatial Interferometer (ISI).  The ISI is a
two-element, heterodyne stellar interferometer operating at discrete
wavelengths in the 9-12~$\micron$ range and is located on Mt. Wilson,
CA.  The telescopes are each mounted within a movable semi-trailer and
together can currently operate at baselines ranging from 4 to 35~m.
Detailed descriptions of the apparatus and recent upgrades can be found
in Bester et al. (1990, 1994).  System calibration
was maintained with the observations of partially resolved 
K giant stars $\alpha$~Tau and $\alpha$~Boo which have no dust known to 
be surrounding them.  Uncalibrated visibilities were found to
drift approximately 15\% from year to year, thus larger systematic
uncertainties may be present when comparing data from different years.
Extended discussions of calibration issues and uncertainties are found in
Danchi et al. (1990, 1994a).
A journal of NML~Cyg observations is provided in Table~1 and contains
the final calibrated visibilities along with error bounds representing
the statistical uncertainty of each visibility measurement
(corresponding to $\sim$15~minutes of on-source integration in most cases).  
Note that the measurement at the longest baseline yielded only an upper
bound to the visibility. The diffraction-limited beam size is $\sim
1\farcs 5$, which does not affect the interpretation of the
visibility curves since previous 10~$\micron$ speckle measurements (Fix
\& Cobb 1988; Dyck \& Benson 1992) have indicated that essentially all
the mid-infrared emission arises from a region 5 times smaller than
this ($\Theta_{\rm{FWHM}} \sim 0 \farcs 3$).

\subsection{Long-term N Band Photometry} 

N~band (10.2~$\micron$) photometry of NML~Cyg from the Maui Space
Surveillance Site (MSSS) (Nishimoto et al. 1995) is presented in
Figure~1.  These data were secured as part of a program monitoring a
number of late-type stars from 1980 to 1992 for the purpose of calibrating
an infrared radiometer system, typically using an N Band equivalent
filter.  The MSSS radiometer, installed on a 1.2~m telescope, consisted of 
a square array of 25~cadmium-doped germanium (Ge:Cd) photoconductive detectors
cooled to 12~K. All reported measurements are
included in Figure~1 and show large fluctuations
($\sim$0.3~mag) from day to day.  Two effects contributed to this large
uncertainty.  Firstly, the MSSS often operated under atmospheric conditions
considered non-photometric by astronomical observatories.  Secondly,
the integration time for each measurement was limited due to the large
number of objects being surveyed each night.  However, the large data
volume available ($\sim$1000 independent measurements over 12~years)
allowed these problems to be mitigated by binning the data into 30 day
windows, with the median of each bin yielding a magnitude estimate
insensitive to the effects of outlying points caused by non-photometric
conditions.  These median data points are depicted on Figure~1
confirm Strecker's (1975) report of a long-term periodicity
of $\sim$1000~days.  The peak-to-trough N~band amplitude is
$\sim$0.5~mag, slightly smaller than the $\sim$0.6~mag variation
reported by Strecker at 3.5~$\micron$ in the early 1970s.  The average
period, as measured from the MSSS photometry, is $\sim$940~days,
although inspection of the light curve reveals that NML~Cyg is not a
regular pulsator.

\begin{figure}
\figurenum{1}
\plotone{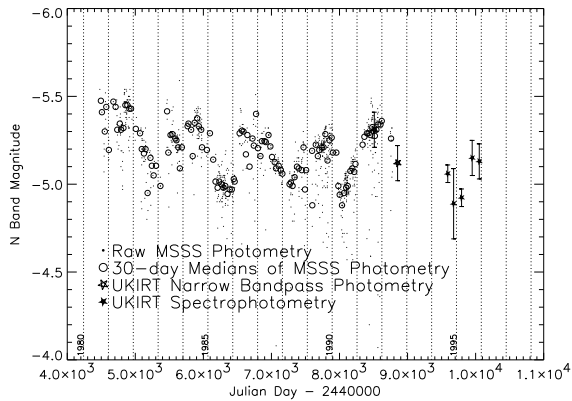}
\caption{The N~band (10.2~$\micron$) magnitude of
NML~Cyg from 1980 to 1996.  The raw data from the Maui
Space Surveillance Site (MSSS) are presented as dots, while the 
circles are the medians of the raw data binned into 30 day windows.
The open stars are N band equivalent flux determinations derived from
narrow band UKIRT observations at 10.5~$\micron$ and 11.5~$\micron$.
The filled stars were derived from 8-13~$\micron$ UKIRT
spectrophotomery.  The flux varies
 $\pm0.25$ magnitudes with an irregular period of $\sim$940 days.}
\end{figure}

\subsection{Mid-infrared Spectrophotometry}

More recent mid-infrared photometry was carried out with the United
Kingdom Infrared Telescope (UKIRT) from 1991 to 1995.  Both
narrowband photometry (using the single-channel bolometer in UKT8) and
8-13~$\micron$ spectrophotometry (using the linear array spectrometer
CGS3) have been obtained with 5$\arcsec$ diameter apertures and
standard chopping and nodding techniques.  Flux calibrations were
derived from observations of $\alpha$~Lyr, $\alpha$~Aur, 
$\alpha$~CMa, and other bright standard stars. In all but one case
the intercomparison of standards on a given night were internally
consistent to within several percent, and hence it is believed that on
those nights the flux level determined for NML~Cyg is accurate to 
$\pm$10\% or better.  On 1994 November 20, which was not photometric, the 
uncertainty
is probably $\pm$20\%.  Details of the spectral shape of NML~Cyg are least 
reliable near 9.7~$\micron$, due to strong telluric absorption by ozone.
Wavelength calibrations were derived from observations of a krypton arc 
lamp in 4th, 5th, and 6th order, and are accurate to $\pm$0.02~$\micron$.

The flux densities at 11.15~$\micron$, the wavelength at which the ISI
observations were obtained, are listed in Table~2, while plots of all
spectra can be found in Figure~2.  The shape of the silicate feature
was nearly constant in all observations despite variations in the total
stellar flux, with the possible exception of the spectra taken near minimum
luminosity.  Based on the MSSS N~band photometry, the 1995 August 20
UKIRT data were chosen to represent the ``mean'' mid-infrared spectrum
of NML~Cyg and appear in Figures~3b and 5b.  All the UKIRT
mid-infrared photometry were converted to equivalent N~band magnitudes
and are plotted on Figure~1 for comparison with
the MSSS photometry.  The two data sets from UKIRT and MSSS 
overlap in time during Fall 1991, and yield internally consistent flux
measurements to within known uncertainties.

\begin{figure}
\figurenum{2}
\plotone{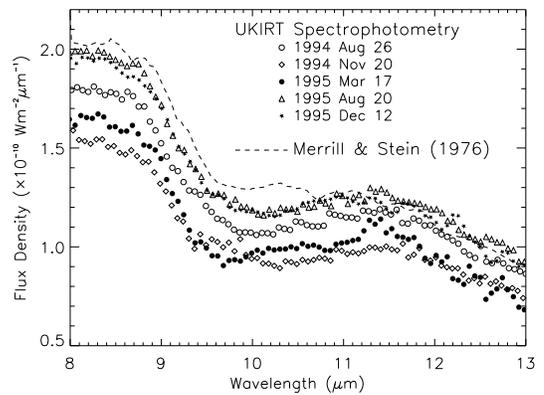}
\caption{The 8-13~$\micron$ spectrum of NML~Cyg at
5~recent epochs (plot symbols) along with a previously published
spectrum from 1976 scaled to the present mean flux level (dashed line).  
The UKIRT spectrophotometric 
data were collected on 1994 Aug 26 (open circles), 1994 Nov 20 (diamonds),
1995 Mar 17 (filled circles), 1995 Aug 20 (triangles), and 1995 Dec 12 
(filled stars). The shape of the silicate feature
changed little during the last pulsational cycle and appears very
similar to a previous measurement taken over 20~years ago.}
\end{figure}

%___________________________________MODELING PROCEDURE_________________
\section{Modeling Procedure}

\subsection{Modeling Code}

One goal of this work is to develop a model of the circumstellar
environment of NML~Cyg which can explain both new and previously
published observations. This requires a method for calculating the
visibility curves and spectra from a given a set of model parameters.
The radiative transfer modeling code used for this purpose was based on
the work of Wolfire \& Cassinelli (1986) and assumes the dust
distribution to be spherically symmetric (see Danchi et al. [1994a] for a
detailed description). Starting with the optical
properties and density distribution of the dust shell, this code
calculates the equilibrium temperature of the dust shell as a function
of the distance from the star, whose spectrum is assumed to be a blackbody.
Subsequent radiative transfer calculations at 67 separate wavelengths
allow the wavelength-dependent visibility curves, the broadband
spectral energy distribution, and the mid-infrared spectrum all to be
computed for comparison with observations.  The dense sampling of the
silicate band is particularly useful for comparison with UKIRT
spectrophotometry.  For a uniform outflow model of the dust shell, a
Sun Sparcstation 10 could perform this calculation within 2 minutes, while more
complex geometries, such as those including multiple dust shells,
generally took $\sim$5~minutes.

The code treats a distribution of dust sizes by calculating the dust
temperature as a function of grain size and dust type (e.g.  dirty
silicates, amorphous carbon, graphite), using the Mathis, Rumpl, \&
Nordsieck (1977) grain size distribution, where the number density $n
\propto a^{-3.5}$, and grain size spanned
$0.01\,\micron$$<a<0.25\,\micron$.  The wavelength-dependent grain
opacity, when averaged over the grain size distribution, was found to
be insensitive to the exact choice of the grain size cutoffs for
wavelengths above 1~$\micron$.  Dust opacities were calculated from the
optical constants assuming spheroidal Mie scattering using a method
developed by Toon \& Ackerman (1981).

For stars having oxygen-rich atmospheres, it is appropriate to use
optical constants for astronomical silicates.  Previous analyses of
oxygen-rich envelopes using the ISI visibility data (e.g. Danchi et al.
1994a,b) have utilized the Draine \& Lee (1984) optical constants for
astronomical silicates when computing the spectrum.  Recent
mid-infrared spectra from UKIRT of late-type stars present the
opportunity to test these constants more rigorously.

The optical constants of Draine \& Lee (DL 1984), Ossenkopf, Henning,
\& Mathis (OHM 1992), and David \& Papoular (DP 1990), were all used in
modeling the dust shell of NML~Cyg.  Calculations of the silicate feature 
using DL constants could not reproduce the observed spectral shape, 
while the more recently determined optical constants of DP and OHM 
were successful at matching observations.  Although OHM and DP constants both
provided similar high-quality fits to the shape of the silicate
feature, OHM dust proved superior in reproducing the shape of the
observed near-infrared spectrum.  The OHM constants in the near-infrared
were determined for laboratory silicates with metallic inclusions (iron and
iron-oxide), leading to a higher opacity (relative to the silicate feature)
than both the DL and DP constants.  We were unable to produce models
using the DL or DP optical constants which matched the excellent fits
resulting from the OHM constants, and hereafter all results presented
are those obtained using the OHM optical constants.  

Radio observations suggest that NML~Cyg is at a distance of approximately
1800~pc (Bowers et al. 1983; Morris \& Jura 1983; Richards et al. 1996)
in an OB association (Cyg~OB2). At this distance, attenuation in the
visible and near-infrared due to intervening interstellar material is
significant, and has been measured by Lee (1970) who found
$E_{\rm{B-V}}$ to be $\sim$1.2 in the direction of Cygnus.  Thus before
comparison with observations, simulated spectra were reddened following
Mathis (1990) with $R_{\rm{V}}=3.1$, yielding a transmission of 25\% at
1~$\micron$, 64$\%$ at 2~$\micron$, and 92\% at 5~$\micron$.  

%%%%%%%%%%%%%%%%%%%%%%%%%%%%%%%%%%%%% Published Observations%%%%%%%%%%%
\subsection{Published Observations}
The new observations of NML~Cyg are important in defining models, but
are unable, by themselves, to adequately constrain the range of
acceptable dust envelope geometries. To this end, a literature search was
conducted in an effort to compile a more comprehensive data set for
comparison with the simulated visibility curves and spectra.  In
fitting all of the compiled data, new observations were given priority
over previously published measurements because secular changes in
NML~Cyg's observables may have occurred, rendering older data less
representative of current conditions.

This section briefly discusses three sets of published data which were
used to guide the modeling process and which appear in our figures:
mid-infrared speckle data, near-infrared speckle data, and
wide-bandpass spectral measurements of NML~Cyg.

The high-resolution 11.15~$\micron$ visibility data were complemented
at low spatial resolution by observations employing mid-infrared speckle
interferometry.  Fix \&
Cobb (1988) first performed this type of measurement in 1985 at the 3~m NASA
Infrared Telescope Facility (IRTF) and Dyck \& Benson (1992) repeated
it using narrowband filters in 1988 and 1989 at the 2.4~m Wyoming
Infrared Observatory.  These results all agree to within uncertainties,
and have been plotted in Figures~3a and~5a, while observational details are
summarized in Table 3.

Sibille et al. (1979), Dyck et al. (1984), and Ridgway et al. (1986)
have all obtained near-infrared speckle interferometric measurements.
The wavelengths, passbands, and dates of all three observations can be
found in Table~3, while the results are plotted in Figures~3c and~5c.
The formal error bars are approximately the same size as the plot
symbols or smaller, although the results from the three papers usually do not
agree with each other to within the quoted uncertainties.  This
disagreement may reflect real changes in the size of NML~Cyg between
the different measurements, or represent inadequate calibration for the
atmospheric seeing.  Ridgway et al. (1986) presented the most recent
and highest resolution set of observations, and these are given greater
weight in the model fitting.  However, the highest resolution points of
this data generally show a trend towards higher visibility.  According
to the authors, the presence of broadband noise proportional to stellar
flux can cause this effect, so following their example, we did not
attempt to fit data beyond approximately 0.75 of the telescope cutoff
(see Figures~1 and~2 of Ridgway et al. [1986]).  However, the entire
data sets are presented in Figures~3c and~5c with a dotted, vertical line
showing the position of the spatial frequency cutoff.

The last data set used in the modeling procedure was comprised of
broadband spectra of NML~Cyg.  Since NML~Cyg has demonstrated
variability in the mid-infrared, it was desirable to find data
containing a broad range of frequency measurements all taken at the
same, recent epoch.  Unfortunately, the most recent spectrum located in
the literature was from Dyck, Lockwood, \& Capps (1974), and the fluxes
taken from this paper, along with fluxes from Johnson, Low, \&
Steinmentz (1965), are given in Figures~3b and~5b.  These groups
collected their measurements over a period of 10 and 20~days
respectively, thus allowing for direct intra-comparison.  The Dyck et al. 
flux measurements were taken near 1971 September 20, while Johnson et al. 
observations were tekn around 1965 May 7.  The error bars represent statistical
measurement uncertainties, but an additional 10\% uncertainty
associated with the absolute calibration has not been plotted.
Strecker's (1975) light curve clearly shows NML~Cyg to be near its
3.5~$\micron$ minimum at the time of the Dyck et al. measurement, and
the absolute flux reported by the two groups agree to within
uncertainties.  The phase of NML~Cyg during the Johnson et al.
measurement can not be extrapolated since NML~Cyg has an irregular
period; however, the 1971 data is systematically lower than the 1965
data, consistent with September 1971 marking a minimum in stellar
luminosity.  The $\lambda=10.2\,\micron$ data points provide an
unexplained exception to this trend, indicating perhaps a larger
uncertainty in the Johnson et al. measurement than
claimed by the authors.

%___________________________________Single SHELL MODEL____________________
\section{Single Shell Models}

Both Rowan-Robinson \& Harris (1983) and Ridgway et al. (1986) have
modeled the circumstellar environment of NML~Cyg as a single dust shell
undergoing uniform outflow.  By also assuming that the dust is
accelerated to a terminal velocity through a distance which is small
compared to the dust shell's full extent, one can 
approximate the density distribution as $\rho\propto
r^{-2}$.  In this model, no dust is presumed to exist between the
stellar surface and the dust shell's inner radius, the location at
which dust condenses out of the gas phase.  Both groups concluded that
the available data could be accurately reproduced by such a model if
the inner radius is chosen to be $\sim 40$ mas from the star. To test
this scenario, a series of models was constructed in which the inner
radius of the dust shell was fixed at 40 mas, while the total optical
depth was varied.

The solid line in Figure~3, corresponding to an 11.15~$\micron$ optical
depth of 2.4, was the result of fitting only to the broadband spectrum.
Note the reasonable agreement with the 11.15~$\micron$ high resolution
visibility data, although the observed mid-infrared size, as measured
by the low resolution speckle observations, is smaller than the model
predicts.  Interestingly, the observed angular diameters at 3.6~$\micron$ and
4.8~$\micron$ show the opposite trend, being larger than the model
simulations.  Table~4 contains a summary of the model parameters.
The dust temperature at the inner radius for this model was $\sim 1200$K, 
consistent with the expected condensation temperature for 
astronomical silicates.

\begin{figure}
\figurenum{3a}
\plotone{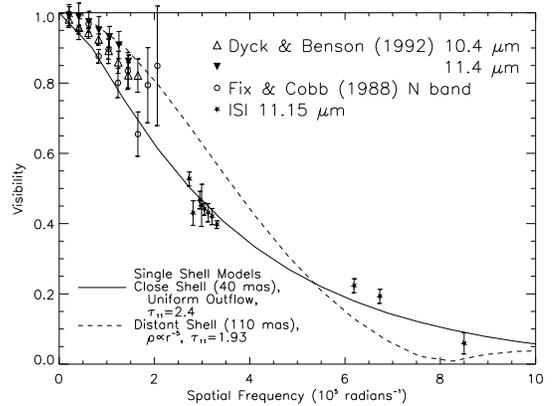}
\caption{The predictions of single shell
models are compared to a set of new and previously published
observations.  The solid line represents results from a single shell,
uniform outflow model of the dust distribution around NML~Cyg, while
the dashed line arises from a model of a single, geometrically thin
shell (see Table~4).  (a) Numerical models are compared with 11~$\micron$
visibility data.  Neither model can fit the short- and long-baseline
data simultaneously.  }
\end{figure}

\begin{figure}
\figurenum{3b}
\plotone{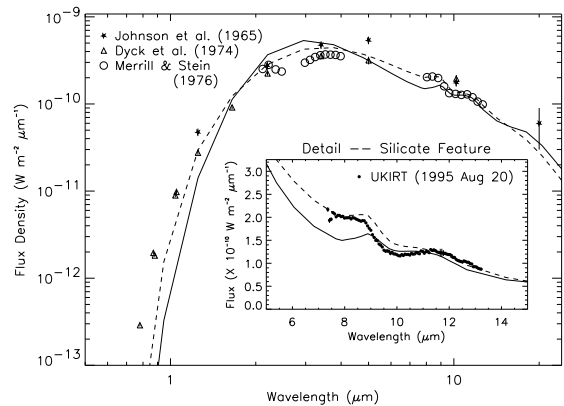}
\caption{The predictions of single shell
models are compared to a set of new and previously published
observations.  The solid line represents results from a single shell,
uniform outflow model of the dust distribution around NML~Cyg, while
the dashed line arises from a model of a single, geometrically thin
shell (see Table~4).  (b) The model spectra are compared to
broadband and mid-infrared spectral measurements.  The geometrically
thin shell model (dashed line) fits these observations much better than
the uniform outflow model (solid line).}
\end{figure}

\begin{figure}
\figurenum{3c}
\plotone{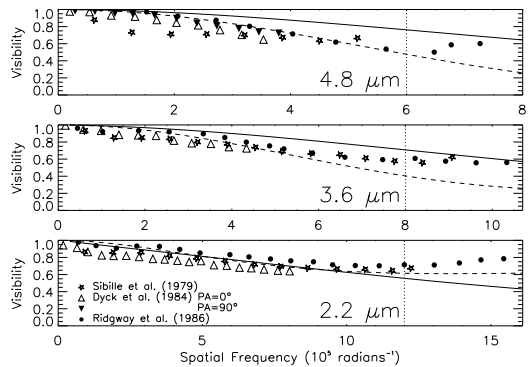}
\caption{The predictions of single shell
models are compared to a set of new and previously published
observations.  The solid line represents results from a single shell,
uniform outflow model of the dust distribution around NML~Cyg, while
the dashed line arises from a model of a single, geometrically thin
shell (see Table~4).  (c) Previously published
near-infrared visibility measurements near 4.8~$\micron$,
3.6~$\micron$, and 2.2~$\micron$ are compared to model calculations.
Error bars are not plotted but are comparable to the size of each data point
or smaller.
Again, the geometrically thin shell model (dashed line) proves superior
in reproducing observations.  The vertical dotted line represents the
spatial frequency cutoff discussed in  \S 3.2.  }
\end{figure}

Further note from Figure 3b that the observed silicate feature is only partly
reproduced by this uniform outflow model.  Although the results were
not plotted, further calculations revealed that a nearly perfect fit to the
silicate feature could be achieved by increasing the optical depth from 2.4
to 3.9.  However, this increase in optical depth severely cut off the broadband
spectrum below 2~$\micron$ and also enhanced the 10~$\micron$ emission
region, worsening the fit to the mid-infrared speckle interferometric 
measurements.

One plausible scenario which accounts for the high optical depth, while
simultaneously avoiding problems with the visible and near-infrared
spectrum, is that there exist holes, or weak spots, in the dust
shell allowing more stellar flux to escape than expected from the
``mean'' optical depth.  A similar effect would be caused by enhanced
forward scattering by the grains.  These possibilities alleviate the
problems of fitting the broadband and near-infrared visibility curves,
and Ridgway et al. (1986) were able to fit their near-infrared
visibility curves by treating the fractional stellar flux as a free
parameter.  However, these scenarios are not consistent with the recent
mid-infrared observations.  The 4.8~$\micron$ optical depth assumed in
the Ridgway et al. (1986) paper is $\sim 4$ times larger than in our
best fitted uniform outflow model (solid line, Figure~3).  Uniform outflow
models with optical depths roughly matching those of Ridgway were
produced for comparison with mid-infrared observations.  Significantly,
no such model was found which could reconcile the relatively small
overall size of the mid-infrared emission region of NML~Cyg, as
indicated by the low resolution speckle measurements, with the large
size implied by the high optical depth of the Ridgway et al. (1986)
model.  Another observation which is difficult to explain with a model
characterized by high optical depth and a large unresolved flux
component is the shape of the silicate feature.  Using OHM optical
constants, a simulation of such an optically thick dust shell
($\rho_{\rm{dust}} \propto r^{-2}$) predicted a significantly deeper
absorption feature centered around 9.7~$\micron$ than was actually
observed.  

The major conclusions drawn by comparing these previous models with
the current data set are:

\begin{enumerate}
\item{Uniform outflow models with dust shell inner radii of 40 mas are
not consistent with 3.6~$\micron$ or 4.8~$\micron$ visibility data.  Without
introducing scenarios such as inhomogeneities in the dust shell or
enhanced forward scattering by grains, the near-infrared visibility
curves are best fitted by a dust shell with a larger inner radius 
since the hottest dust contributes most to the resolved
component of the near-infrared visibility curves.}

\item{The Dyck \& Benson and Fix \& Cobb 11~$\micron$ visibility data both
show NML~Cyg's dust shell to be much ``smaller'' 
than expected for a $\rho \propto
r^{-2}$ density distribution with sufficient optical depth to move the
silicate feature into absorption.  The long-baseline 11.15~$\micron$ data
alone could be adequately fitted by a uniform outflow model, but such a
model can not fit the 11~$\micron$ speckle data
simultaneously.}

\end{enumerate}

In response to the major shortcomings of the uniform outflow models, a
series of single shell models were produced in which the dust density
behaved as $\rho \propto r^{-5}$.  Both the inner radius and optical
depth were varied in an attempt to fit the broadest possible set of
observations.  The choice of dust shell geometry, $\rho \propto
r^{-5}$, was based on the results of the previous modeling, where it
was observed that the near-infrared emission arose almost entirely in a
thin region around the inner radius of the dust shell, because only the
hottest of the dust contributes significantly to the visibility curve.
This implied that a density distribution falling more steeply than
$r^{-2}$ would have little effect on the near-infrared spectrum or
visibility curves, but would have the desired effect of shrinking the
mid-infrared size, which was dominated by the emission from cool dust
(T $\lesssim 600$~K) far from the star.

Predictions from the $\rho \propto r^{-5}$ dust shell model which fit
the most diverse set of observations can be found as the dashed line in
Figure~3, while a summary of the model parameters has been placed in
Table~4. The best fitted model consisted of a single shell with an inner
radius of 110~mas and an 11.15~$\micron$ optical depth of 1.93. This
model fits most of the observations, including the near-infrared
visibility data, mid-infrared speckle observations, the broadband
spectrum (above 1~$\micron$), and the mid-infrared silicate feature.
Similar quality results were also achieved for steeper density laws,
e.g. $\rho \propto r^{-7}$.  Hence by allowing the dust shell geometry
to deviate from uniform outflow, it was possible to reproduce a much
broader range of observations.  In fact, this single shell model can
explain essentially all the observations, with the sole exception of
the long-baseline 11.15~$\micron$ visibility data introduced earlier in
this paper.

%_________________MEM RECONSTRUCTION OF MID-INFRARED__________________
\section{Maximum Entropy Reconstruction of the Mid-Infrared Brightness
Distribution}

The Maximum Entropy Method (MEM), based on the algorithm of Wilczek \&
Drapatz (1985), was used to reconstruct a circularly symmetric
brightness distribution.  The 11.4~$\micron$ speckle data of Dyck \&
Benson (1992) were chosen to represent the low resolution data, while
the high resolution information was provided by the ISI.  Figure~4
contains the MEM reconstruction of the radial profile of NML~Cyg,
where the brightness distributions plotted are weighted by the
radius, so that the area under each curve is proportional to the total
monochromatic flux.  The MEM fit to the 11~$\micron$ visibility points
is statistically adequate ($\chi^2 \sim $number of degrees of freedom),
and is plotted in the inset plot of Figure 4.

\begin{figure}
\figurenum{4}
\plotone{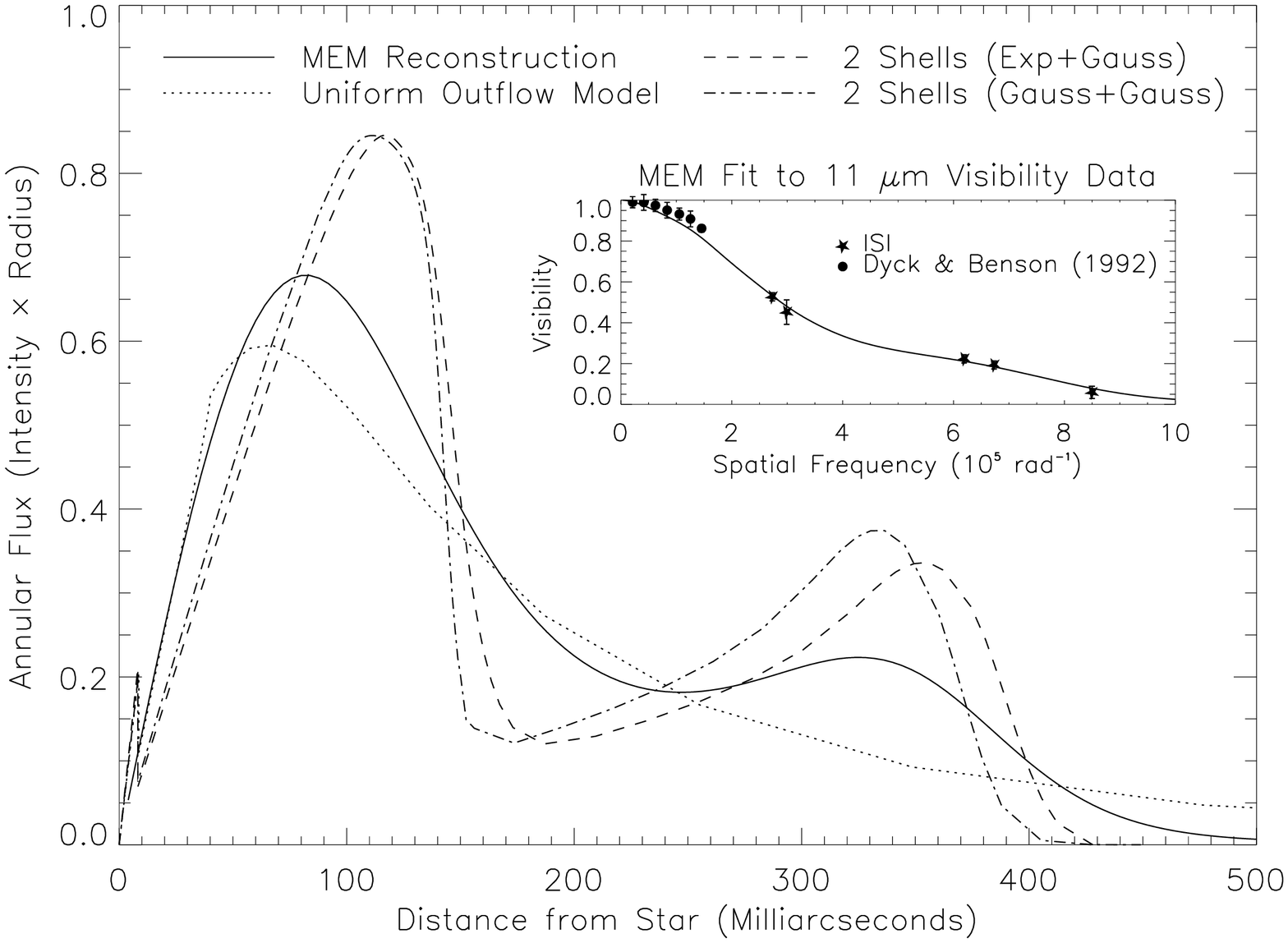}
\caption{The~11 $\micron$ annular flux
from the Maximum Entropy Method (MEM) reconstruction of NML~Cyg is
displayed as a solid line, while the visibility data are shown in the inset
plot.  Only the visibility data taken before 1996 were utilized in the
MEM reconstruction.
The suggestion of a two-shell geometry is apparent when the MEM
result is compared with the results from a uniform outflow model
(dotted line).  In addition, the calculated annular flux of
various dust shell models are included for comparison with the MEM
results.  The dashed line is from a double
shell model where the inner shell dust density is mathematically expressed
as an exponential, while the dash-dotted line represents a double shell model
in which the inner shell density is described by a radial Gaussian.  In both
double shell models, the outer shell is described by a radial Gaussian
(see Table~4 for model parameters).  Note that the annular flux
(brightness distribution weighted by the radius) is displayed
to make more prominent low surface brightness
features which provide significant total flux.}
\end{figure}

Despite the MEM algorithm's attempt to spread the flux out as much as
possible in image space while adequately fitting the visibility data, 
Figure~4 shows evidence for a double shell structure.  For comparison, the 
best-fit
uniform outflow model from the previous section (solid line, Figure~3)
is included on the figure as the dotted line.  The MEM result suggests
a two component dust shell geometry with the closest shell appearing
approximately 100~mas from the star, similar to the location of the
shell in the $\rho_{\rm{dust}} \propto r^{-5}$ model discussed
previously (dashed line, Figure~3).

%_______________________________DOUBLE SHELL MODELS___________________
\section{Double Shell Models}

With the MEM results it was possible to construct a realistic, two
component model which fitted nearly all the observations.
Schematically, the two shells consist of one geometrically thin shell
located approximately 100~mas from the star (similar to the $\rho
\propto r^{-5}$ shell discussed previously), and a distant, colder
shell approximately 300~mas from the star.  The first shell alone has
already been shown to reproduce most of the observations, with the
exception of the 11.15~$\micron$ observations.   The addition of some
dust far from the star should have a minimal effect on the
near-infrared observations because the low temperature dust contributes
little to the near-infrared output.  However, the MEM reconstruction
indicates that such a dust shell can be used to reproduce the 
high-resolution 11.15~$\micron$ visibility data.

The parameterization of the inner dust shell was inspired by David \&
Papoular (1992).  They determined that the dust density distribution
around many optically thick IRAS sources could not be modeled with a simple
$r^{-2}$ density law, but rather required a geometrically thin,
exponential shell near to the star and a uniform density ring far from
the star in order to reproduce the observed IRAS colors and deep
silicate absorption feature.  NML~Cyg does not have such a deep
silicate feature, but neither is it optically thin at 10~$\micron$.

For convenience, the David \& Papoular type distribution was used, with
the inner dust shell being described by $\rho \propto
exp(-\frac{r-R_{\rm{inner}}}{l_{\rm{scale}}})$ for $r >
R_{\rm{inner}}$, where $l_{\rm{scale}}$ is the 1/e scale height. The
density in the outer dust shell had a simple Gaussian radial
dependence.  The functional form of the dust shell density distribution
can be found in Table~4.  The scale height of the double shell model
used in this analysis was smaller than those used by David \& Papoular
(1992) because of their numerical code limitations.  However, radiative
transfer calculations of dust distributions using two Gaussian shells
were also performed in order to test the sensitivity of the shell
parameters to the chosen density distribution.  These results are
summarized in Table~4 and will be discussed in the next section.  Note
that the outer shell found in this paper differs significantly in character
from that postulated by David \& Papoular (1992).  The uniform density shell
used in that work was located much farther away from the star than the
shell used here, and was primarily useful for fitting the
far-infrared flux and the deep absorption found in the silicate
feature, two observations not particularly relevant to this study.

\begin{figure}
\figurenum{5a}
\plotone{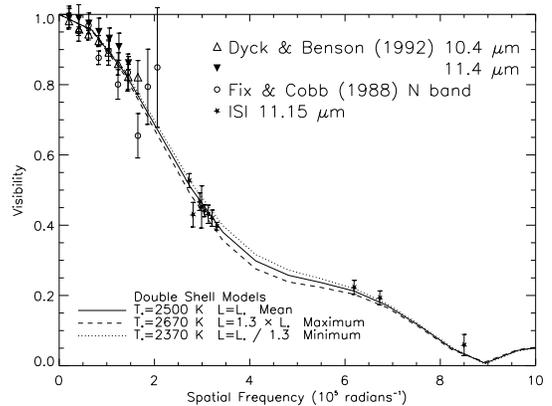}
\caption{The predictions from a double shell model
are presented along with both recent and previously published 
observations.  These results represent NML~Cyg at mean luminosity
(solid line), maximum luminosity (dotted line), and minimum luminosity 
(dashed line). 
(a) Model calculations are compared to observed 11~$\micron$
visibility data.  Changes in luminosity do not appear to have significant
effects on the 11~$\micron$ visibility.}
\end{figure}

\begin{figure}
\figurenum{5b}
\plotone{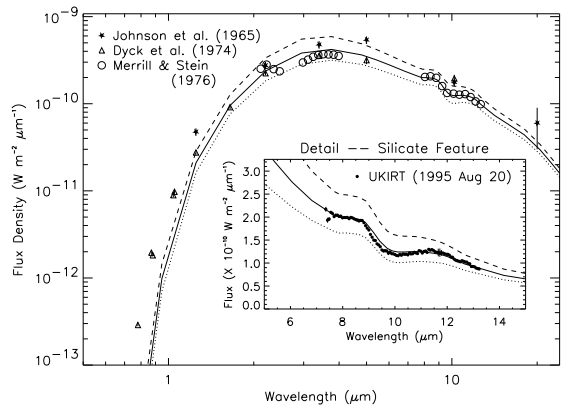}
\caption{The predictions from a double shell model
are presented along with both recent and previously published 
observations.  These results represent NML~Cyg at mean luminosity
(solid line), maximum luminosity (dotted line), and minimum luminosity 
(dashed line).  
(b) The model spectra are compared to
broadband and mid-infrared spectral measurements.  Note the broadband
spectrum is not well-fitted below 1.5~$\micron$.  Ossenkopf et al. (1992)
dust constants help provide the excellent fit to the silicate feature.}
\end{figure}

\begin{figure}
\figurenum{5c}
\plotone{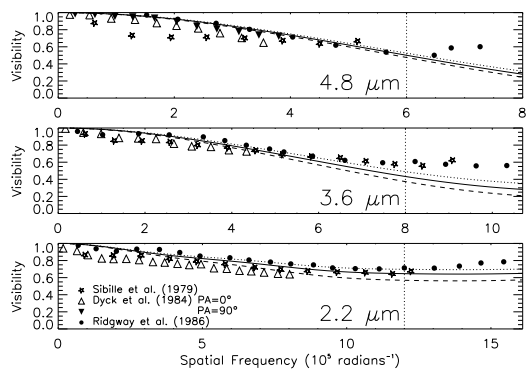}
\caption{The predictions from a double shell model
are presented along with both recent and previously published 
observations.  These results represent NML~Cyg at mean luminosity
(solid line), maximum luminosity (dotted line), and minimum luminosity 
(dashed line). 
(c) Previously published
near-infrared visibility measurements near 4.8~$\micron$,
3.6~$\micron$, and 2.2~$\micron$ are compared to model calculations.
The vertical dotted line represents the
spatial frequency cutoff discussed in  \S 3.2. Although fitting the
observations well below this cutoff, the models predict high resolution
visibility points systematically lower than published observations.
These high-resolution values were not fitted in the original publication, 
and are presumed to be somewhat unreliable.}
\end{figure}

Using the MEM reconstruction as a guide, it was straightforward to
determine a set of two-shell parameters which fit the 11~$\micron$
visibility curve (see Table~4), although such parameters were not
unique.  In fitting the bump in the 11~$\micron$ visibility curve, the
possible values of the shell radii and relative optical depths were
constrained.  The silicate feature was then fitted by adjusting the
overall optical depth while keeping the ratio of optical depths roughly
the same.  As can be seen in Figure~5b, this model provides an excellent
fit to the silicate feature. The shape of the silicate feature was
found to be relatively insensitive to the total optical depth in this
region of parameter space, and the 11.15~$\micron$ optical depth
of the inner shell could be adjusted from 1.8 to 2.5 and still maintain
adequate fits to the mid-infrared spectrum and 11~$\micron$ visibility
data.

Advantage was taken of this parameter flexibility in fitting the
broadband spectrum.  By decreasing the inner shell's optical depth to
1.8, the previous fits to the silicate feature and 11.15~$\micron$
visibility data were maintained while additionally fitting the
broadband spectrum for wavelengths longer than $\sim$1.5~$\micron$ (see
Figure~5b).  Possible explanations for this model's failure to fit the
spectrum for wavelengths shorter than 1.5~$\micron$ are discussed
below.  Note that this broadband spectrum fit is superior to that
achieved by the uniform outflow model in Figure~3b.

Finally, the trial model was compared to the visibility data in the
near-infrared.  There was little room left for tuning the parameters,
because the fits to the broadband spectrum and the mid-infrared data
strictly constrained the possible values for the shell inner radii and
thicknesses. Furthermore, the overall optical depth of the shells was
constrained by the requirement to fit the near-infrared part of the
broadband spectrum.  The remaining parameter space was exploited in order
to optimize the fit to the near-infrared visibility data and the fits
for the best double shell model are shown Figure~5c.
Clearly the data are fitted adequately by this model, which does not
require any arbitrary tuning of optical depths or contributions from an
unresolved point source.  Residual misfits to the data are as
likely to arise from uncertainties in the dust constants, variations in
luminosity, or systematic errors in the measurements themselves, as
they are from an incorrect model of the dust shell geometry.  Probably
all of these factors contribute to the residual misfits.

The best fitted double shell model was characterized by an inner shell
which began at 120~mas with a scale height of 10~mas.  The outer,
Gaussian shell was centered around 370~mas with a standard deviation of
20~mas, although this model was not very sensitive to parameters of the
outer shell.  The respective optical depths of the inner and outer
shell at 11.15~$\micron$ were 1.8 and 0.34.  Given that the density
distribution has the correct functional form, the parameters of the
inner shell have an approximate uncertainty of 10\%, while the outer
shell's uncertainties are larger ($\sim 20$\%) because it was not
possible to discriminate between a thin shell at 380~mas and a thicker
one at 350~mas.

%_____________________________________DISCUSSION_________________________
\section{Discussion}

The two-shell model was tested to see what changes in the simulated
spectrum and visibility curves arose due to varying the luminosity and
temperature of the star.  The temperature of NML~Cyg was modeled as
2500~K by Rowan-Robinson \& Harris (1983), while Ridgway et al. (1986)
used a stellar temperature of 3250~K, appropriate for an M6III star.
For a fixed stellar luminosity, changing the effective temperature
mainly affects the near-infrared visibility curves.  In order to
examine the model's sensitivity to the stellar $T_{\rm_{eff}}$,
simulations were performed with $T_{\rm_{eff}}=2500\,\rm{K}$ and
3000~K, while holding the luminosity constant.  We found that there was
virtually no effect on the fit to the 11~$\micron$ visibility data and
the broadband spectrum, and that the most significant departure was
that the amount of unresolved stellar flux at 2.2~$\micron$
increased by $\sim$5\% for the lower stellar temperature.  As the
near-infrared data are not internally consistent to this level, this
effect is not considered further.

Since the data collected on NML~Cyg were not all from the same epoch,
the double shell model discussed above was computed with different
stellar luminosities in order to take into account the known variability
of this star.  Plotted in Figure~5 are model curves for various
luminosities of the source, consistent with the maximum and minimum
10~$\micron$ flux density from MSSS data (see Figure~1).  The change in
luminosity was accomplished by fixing the stellar radius and changing
only the temperature, which would have exaggerated any temperature
effects on the model visibility curves and spectrum. The solid line in
Figure~5 represents roughly the mean flux and was used to fit the
mid-infrared spectrum taken at UKIRT on 1995 August 20, while the other
curves represent models at the maximum and minimum luminosities.
Fortunately, the change in luminosity had only a small effect on the
mid-infrared visibility predictions, although the 2.2~$\micron$
visibility curves were more sensitive.  This sensitivity to the stellar
luminosity might explain some of the discrepancies in the near-infrared
speckle data.  Further inconsistencies between NML~Cyg observations
taken at different times could occur if the dust condensation radius
changed significantly during a luminosity cycle.  There is currently
not enough information to properly model this, however this effect has
been seen on other stars such as $o$ Cet (Lopez et al. 1996) and
IRC+10216 (Danchi et al.  1994a).

Another check was made to determine the sensitivity of the two-shell
model to the detailed
geometry of the dust shell.  A model similar to the previously 
discussed two-shell model was created which used
an inner Gaussian shell instead of an inner exponential shell.  This
resulted in only slightly different values for the inner radii and the
optical depths, and the best-fit parameters of both double shell models
are located in Table~4.  
Figure~4 compares the 11~$\micron$ brightness
distribution of both two-shell models with the MEM and uniform outflow
results.  The 11~$\micron$ visibility data predictions from both double
shell models can be seen to be nearly identical.  

Although successful at explaining a large body of observations, neither
single nor double shell models were able to fit the broadband spectrum
adequately below 1.5~$\micron$. This is probably due to some
combination of the following effects.  (1) Interstellar reddening
corrections for NML~Cyg, which
are only very important below 2~$\micron$, are not well
known. (2) The relative strengths of the near-infrared dust opacity to
the mid-infrared opacity may not be correct (see discussion by OHM
[1992] detailing the determination of the near-infrared optical
constants). (3) The spectrum could have changed in the last 20 years
(see discussion below). (4) Because of the large extinction due to dust
in the visible and near-infrared, inhomogeneities in the dust shell
could allow additional stellar flux to skew the spectrum.  This latter
scenario requires $\sim$10\% of the 2.2~$\micron$ flux to be stellar
light unreddened by the dust for the spectrum below 1.5~$\micron$ to be
adequately fit. Determining which of these scenarios apply will require
long-baseline, near-infrared visibility measurements to determine the
fraction of light which is stellar.

It is important to emphasize that these models all assume spherical
symmetry.  However, if non-symmetric models are considered, then the
outer shell of the double shell models could be interpreted as a blob
or clump of emission with a characteristic distance of 370 mas from
the stellar surface.  The position angles of the ISI measurements given in
Table~1 could be used to investigate possible asymmetric dust
distributions, although present data does not encompass sufficient
sampling of the Fourier plane to justify full two-dimensional modeling.
However, it is interesting to note that the inner and outer shells have
similar masses, to within a factor of 2.  This suggests that NML~Cyg
may experience episodic mass ejections, as seen in other supergiants,
e.g.~$\alpha$~Ori and $\alpha$~Sco.  It is expected that the outer
shell is beyond the acceleration zone and has attained terminal
velocity.  The known CO outflow velocity of $\sim$25~km/s allows us to
estimate the time between the hypothetical ejections as
$\sim$80~years.

To test the hypothesis of episodic mass ejection, a third shell of the
same mass was placed around NML~Cyg at various radii to see the
effect on the 11~$\micron$ visibility curves.  These radiative transfer
results indicated that the small angular size seen by 11~$\micron$ speckle
observers rule out another such shell within $\sim$1$\arcsec$.  This
result is consistent with the Herbig \& Lorre (1974) image at 0.8~$\micron$ 
which revealed no nebulosity further than 375~mas from NML~Cyg.

The dynamics of the inner shell are not easy to discern. One might
expect the inner shell to have condensed out of the gas phase
approximately 40~mas from the star, where the temperature was
$\sim$1200~K.  The assumption that the dust shell was subsequently
accelerated to the same velocity as the outflowing CO gas
($\sim$25~km/s) would predict that the inner shell was $\sim$60~mas from
the star in 1974, when Merrill \& Stein (1976) measured the
2-13~$\micron$ spectrum of NML~Cyg.  Their measurements were averaged
over a significant part of the cycle, and hence only a relative
spectrum was given.  To permit comparison with the present
simulations, the spectrum was scaled by a constant factor in order to
match the UKIRT observations of 1995 August 20.  The scaled results of
Merrill \& Stein (1976) have been placed on Figures~2, 3b, and 5b, and
are consistent with more recent observations.  Note the spectral feature seen
in the Merrill \& Stein spectrum near 2~$\micron$ was not included in
the numerical computations and does not appear in the model curves.
A dust shell model with
the inner shell placed approximately 60~mas from the star (but
retaining the same functional form and total dust mass) was generated
to compare with the silicate feature observed by Merrill \& Stein in
the early 1970s. The radiative transfer calculation of such a model
showed drastic changes in the spectrum and silicate feature
inconsistent with spectra given in Figure~2. This indicates that the
inner shell has had a velocity substantially lower than 25~km/s, which is
consistent with H$_2$O maser observations discussed below.

The constancy of the silicate feature over the last 20~years suggests
other mass-loss scenarios.  David \& Papoular (1992) envision the
optically thick shell as being too heavy to be accelerated away by
radiation pressure.  Such a shell would presumably continue to collect
mass at approximately the same distance from the star until some
unusual activity of the star causes a mass ejection.  Alternatively,
the dust may have formed closer to the star, but is still accelerating
and has a smaller velocity than the terminal flow.  Considering the
uncertainties in the model, the mid-infrared spectra taken 20~years
apart are too similar to unambiguously prefer one of these scenarios
over another.

Richards et al. (1996) recently presented new measurements of the
22~GHz H$_2$O maser emission around NML~Cyg.  These authors found an
irregular ring of masers approximately 200~mas across and a pair of
features further away, $\sim$600~mas from each other, on either side of
the star.  The double shell models used in this paper were generated
before the Richards et al. (1996) observations were released, and it is
interesting to note the similarity in the location of the masers and
the twin dust shells derived independently.  In addition, proper
motions of the distant masers themselves indicate that the masers have
velocities consistent with the observed CO gas (19$\pm$4~km/s assuming
a distance of 2000~pc).  The Doppler velocities of the inner shell of
masers indicate that the circumstellar material is gravitationally
bound when it enters the H$_2$O maser region (v$\sim$15~km/s at
50~mas), but leaves the maser zone with a velocity larger than escape velocity
(v$\sim$25~km/s at 240~mas).  Richards et al. (1996)
conclude the inner shell of H$_2$O masers is still in the acceleration
regime.

%___________________________________CONCLUSIONS________________________
\section{Conclusions}

In summary, uniform outflow models of the dust envelope of NML~Cyg could
not be simultaneously reconciled with both near- and mid-infrared
observations.  A single, geometrically thin shell of intermediate
optical depth 
can reproduce most of the observations except for the long-baseline, mid-infrared
interferometric data.  An additional shell of dust was required to fit these
high-resolution observations, which are solely sensitive to 
the relatively low-temperature dust existing hundreds of
milli-arcseconds from the stellar surface.  Dust shell morphology
is important in understanding the possible mechanisms which drive the
high mass loss rates observed around red giants.  The Ossenkopf, Henning, \&
Mathis (1992) optical constants were found to be superior to those of
David \& Papoular (1990) and of Draine \& Lee (1984) in providing model
fits to both the detailed shape of the 9.7~$\micron$ silicate feature and
most of the broadband spectrum.  The excess flux
observed below 1.5~$\micron$ has several possible explanations, although
the exact cause is not known.

The double shell models which best fit the observations consist of an
inner shell approximately 125 mas from the star with an 11~$\micron$
optical depth of $\sim$1.9 and characteristic thickness of $\sim$15
mas.  The outer shell was found to reside three times further away with an 
11~$\micron$ optical depth of $\sim$0.33, and contains roughly the same mass as
the inner shell.  Although the models were relatively insensitive to the particular
dust shell geometry, the exact values of these parameters could change by
$\sim$10\%  depending on the functional form of the dust shell density
distributions (see Table~4).

Deviations from spherical symmetry will necessarily change the meaning
of the above fits.  A dense clump of material 350~mas from the star
would be interpreted here as a shell and would cause the inner shell to
be modeled thinner than in fact it is.  These effects can be important, and
confidence in the interpretation should wait until full (u,v) coverage
is obtained.

Coordination of high-resolution observations in the near- and
mid-infrared would place the strongest constraints on models of the dust
shell
geometry and optical constants.  A factor of 2 improvement in
resolution will allow for the near-infrared observations to fully
resolve the dust and thus yield a measurement of the fraction of 
stellar flux escaping the envelopes directly.  
These data should be available in the 
future as a new generation of multi-element infrared interferometers
become operational.  Dynamical simulations of red giant atmospheres will
be required to fully interpret such new observational data.  The
U.C.  Berkeley Infrared Spatial Interferometer will continue to observe
NML~Cyg on a regular basis.

%___________________________________ACKNOWLEDGEMENTS___________________
\acknowledgments
{ 
We thank W.-H. Tham and C.D. Matzner for enlightening discussions.  N.
Turner and J. Graham were both helpful in obtaining the most recent
reddening curves, and V. Ossenkopf and P. David graciously provided
computer-readable versions of their optical constants.  Long-baseline
interferometry in the mid-infrared at U.C. Berkeley is supported by the
National Science Foundation (Grants AST-9321384, AST-9321289, \&
AST-9500525) and by the Office of Naval Research (OCNR
N00014-89-J-1583  \& FDN0014-96-1-0737). M. J. was supported under the
auspices of the U.S. Department of Energy by Lawrence Livermore National
Laboratory under Contract W-7405-ENG-48. E. L. was supported by a
National Science Foundation graduate fellowship during part of this work.  
The United Kingdom Infrared Telescope is operated by the Joint Astronomy
Centre on behalf of the U.K. Particle Physics and Research Council.
} \pagebreak

%___________________________________BIBLIOGRAPHY_______________________
%\begin{thebibliography}{} 

%%%%%%%%%%%%%%%%%%%%%%%%%%%%%%%%%%%%
%% Next goes the Tables here      %%
%%%%%%%%%%%%%%%%%%%%%%%%%%%%%%%%%%%%

\clearpage
%%%%%%%%%%%%%%%%%%%%%%%%%%%%%%%%%%%%%%%%%%%%%%%%%%%%%%%%%%%%%%%%%%
% Table One contains the ISI Journal information                 %
%%%%%%%%%%%%%%%%%%%%%%%%%%%%%%%%%%%%%%%%%%%%%%%%%%%%%%%%%%%%%%%%%%
\begin{deluxetable}{cccc}
\tablewidth{0pt}
\tablecaption{ISI Journal of Observations}
\tablehead{
\colhead{Date} & \colhead{Spatial Frequency} & \colhead{Visibility} & 
\colhead{Position Angle\tablenotemark{a}}  \nl
     & \colhead{($10^{5}$ $\rm{rad}^{-1}$)}& & \colhead{(Degrees)}   \nl
}
\startdata
1994 Sep 22 & 2.73 & $0.53\pm 0.02$ & $50\pm 10$   \nl
1996 Oct 10 & 2.81 & $0.43\pm 0.03$ & $50\pm 3 $   \nl
1996 Sep 23, Oct 10 & 2.95 & $0.47\pm 0.03$ & $56\pm 2$  \nl
1994 Sep 15 & 2.99 & $0.45\pm 0.06$ & $58\pm 2$    \nl
1996 Sep 23, Oct 10 & 3.05 & $0.44\pm 0.02$ & $60\pm2$  \nl
1996 Sep 23, Oct 7,10 & 3.13 & $0.43\pm 0.03$ & $63\pm1$ \nl
1996 Oct 7,10 & 3.21 & $0.42\pm 0.02$ & $66\pm 2$ \nl
1996 Oct 7,10 & 3.31 & $0.40\pm 0.01$ & $70\pm 2$ \nl
1993 Sep 30 & 6.19 & $0.22\pm 0.02$ & $93\pm 3$    \\
1993 Sep 30 & 6.73 & $0.19\pm 0.02$ & $97\pm 3$    \nl
1993 Jul 27 & 8.50 & $0.059\pm 0.030$ & $140\pm 5$   \\ 
1994 Jul 8,9,14 & 27.5 & $<0.07$ & $ 130\pm 20 $   \\
\enddata
\tablenotetext{a}{Position angle is measured in degrees East of North}
\end{deluxetable}

%%%%%%%%%%%%%%%%%%%%%%%%%%%%%%%%%%%%%%%%%%%%%%%%%%%%%%%%%%%%%%%%%%
% Table Two contains the Geballe information                     %
%%%%%%%%%%%%%%%%%%%%%%%%%%%%%%%%%%%%%%%%%%%%%%%%%%%%%%%%%%%%%%%%%%
\begin{deluxetable}{ccc}
\tablewidth{0pt}
\tablecaption{Journal of UKIRT Photometry at 11.15 $\micron$}
\tablehead{
\colhead{Date}  &   \colhead{Flux}           & \colhead{Flux}              \\
      &  \colhead{($10^{-12}$ W{m}$^{-2}{\micron}^{-1}$)} 
& \colhead{($10^{3}$ Jy) }       \\ }
\startdata
1991 Sep 12 & $141\pm 14$& $5.9\pm 0.6$ \\
1992 Aug 19 & $122\pm 12$& $5.1\pm 0.5$ \\ 
1994 Aug 26 & $114\pm 6$& $4.74\pm 0.24$ \\
1994 Nov 20 &  $97\pm 19$& $4.0\pm 0.8$ \\
1995 Mar 17 & $104\pm 5$& $4.30\pm 0.22$ \\
1995 Aug 08 & $125\pm 13$& $5.2\pm 0.5$ \\
1995 Dec 02 & $122\pm 12$& $5.0\pm 0.5$ \\
\enddata
\end{deluxetable}

%%%%%%%%%%%%%%%%%%%%%%%%%%%%%%%%%%%%%%%%%%%%%%%%%%%%%%%%%%%%%%%%%%
% Table Three contains the Speckle Journal information           %
%%%%%%%%%%%%%%%%%%%%%%%%%%%%%%%%%%%%%%%%%%%%%%%%%%%%%%%%%%%%%%%%%%

\begin{deluxetable}{ccccl}
\tablewidth{0pt}
\tablecaption{Journal of Speckle Observations}
\tablehead{
\colhead{Date}  & \colhead{ $\lambda$} & \colhead{$ \Delta \lambda$} & 
\colhead{Position Angle } & \colhead{Reference} \\
      &  \colhead{($\micron$) } & \colhead{($\micron$) }         & \colhead{(Degrees)}       &    \\ }
\startdata
1978 & 2.2 & 0.09 & 120  & Sibille et al. 1979 \\
May 20	  & 3.5 & 0.57 & 120  &  \\
	  & 4.7 & 0.21 & 120  & \\ \nl
1981-1983 &  2.2 & 0.4   &  0   &  Dyck et al. 1984  \\
 &  3.8    & 0.6   &  0  & \\
      &  4.8       & 0.5   &  0,90  &  \\ \nl
1983 & 2.2        & 0.1&  0,90  & Ridgway et al. 1986 \\
Jun 29-30  & 3.45       & 0.57& 0,45,90  & \\
           & 4.95       & 0.5 & 0  & \\ \nl
1985 & 10	& 5.0 & 0 & Fix \& Cobb 1988 \\ 
Jul 21-22 & & & & \\ \nl
1988-1989 & 10.4           & 1.0   &  0    & Dyck \& Benson 1992 \\
  & 11.4       & 1.0   &  0    & \\ 
\enddata
\end{deluxetable}

%%%%%%%%%%%%%%%%%%%%%%%%%%%%%%%%%%%%%%%%%%%%%%%%%%%%%%%%%%%%%%%%%
% Table Four contains the Model Parameters                      %
%%%%%%%%%%%%%%%%%%%%%%%%%%%%%%%%%%%%%%%%%%%%%%%%%%%%%%%%%%%%%%%%%

\begin{deluxetable}{lccllll}
\small
\tablewidth{0pt}
\tablecaption{Model Parameters}
\tablehead{
\colhead{Model Description} & \colhead{$R_\ast$} & \colhead{$T_\ast$} & 
\colhead{Shell One}  & \colhead{Shell Two} & \colhead{Comments} \\
		  & \colhead{(mas)}  & \colhead{ (K)} &\colhead{Parameters} & 
\colhead{Parameters} &  \\ }
\startdata
Single Shell:	  & 8.6 & 2500& $\rho \propto r^{-2}$ &  & Fits 11 $\micron$ visibility.\\
Uniform Outflow   & &&$R_{\rm{inner}}=40$ mas & & Poor fit to 4.8 $\micron$\\
		  & &&$\tau_{11}=2.4$ & & speckle data and  \\ 
		  & &&                & & broadband spectrum. \\  \nl

Single Shell: & 8.6 & 2500 &$\rho \propto r^{-5}$ & & Fits all data except for\\
Geometrically Thin&&  & $R_{\rm{inner}}=110$ mas & & long-baseline 11 $\micron$\\
	&&& $\tau_{11}=1.93$  & & interferometry \\ \nl

Double Shell: & 8.2 & 2500 & $\rho \propto e^{-(r-R_1)/l}$&  $\rho \propto e^{-{\left[ \frac{(r-R_{2})}{\sigma_2}\right]}^2}$ & Fits all data except \\
Exponential \&  && & $R_1=$120 mas & $R_2=$370 mas & spectrum below 2 $\micron$ \\
Gaussian &&&  $l=10$ mas & $ \sigma_2=20$ mas &  \\
 &&&$\tau_{11}=1.8 $&$\tau_{11}=0.34$ & & \\ \nl

Double Shell: & 8.2 & 2500 &$\rho \propto e^{-{\left[{\frac{(r-R_{1})}{\sigma_1}}\right]}^2}$ &  $\rho \propto e^{-{\left[\frac{(r-R_{2})}{\sigma_2}\right]}^2}$ & Fits all data except \\
Gaussian  \& && & $R_1=$125 mas & $R_2=$350 mas & spectrum
 below 2 $\micron$ \\
Gaussian &&&   $ \sigma_1=10$ mas  & $ \sigma_2=20$ mas &  \\

        &&&$\tau_{11}=2.0 $&$\tau_{11}=0.33$ & & \\ \nl

\enddata
\end{deluxetable}

\end{document}